\documentclass[12pt,preprint]{aastex}
\newcommand{\etal}{{\it et al.}}
\newcommand{\beq}{\begin{equation}}
\newcommand{\eeq}{\end{equation}}
\newcommand{\ben}{\begin{eqnarray}}
\newcommand{\een}{\end{eqnarray}}
\newcommand{\br}{{\bf r}}

\newcommand{\xp}{x^{\prime}}
\newcommand{\yp}{y^{\prime}}
\newcommand{\zp}{z^{\prime}}
\newcommand{\tPhi}{\tilde{\Phi}}
\newcommand{\tSigma}{\tilde{\Sigma}}
\newcommand{\ep}{\epsilon}
\newcommand{\lam}{\lambda}
\newcommand{\tlam}{\tilde{\lambda}}

\begin{document}
\title{Reconstructing Three-dimensional Structure
of Underlying Triaxial Dark Halos From X-ray and Sunyaev--Zel'dovich 
Effect Observations of Galaxy Clusters}
\author{\sc Jounghun Lee and Yasushi Suto}
\affil{Department of Physics, The University of Tokyo, 
 Tokyo 113-0033, Japan}
\email{lee@utap.phys.s.u-tokyo.ac.jp, suto@phys.s.u-tokyo.ac.jp}
\received{2003 May 29}
\accepted{2003 ???}
\begin{abstract} 
While the use of galaxy clusters as {\it tools} to probe cosmology is
established, their conventional description still relies on the
spherical and/or isothermal models that were proposed more than 20 years
ago. We present, instead, a deprojection method to extract their
intrinsic properties from X-ray and Sunyaev--Zel'dovich effect
observations in order to improve our understanding of cluster physics.
First we develop a theoretical model for the intra-cluster gas in 
hydrostatic equilibrium in a triaxial dark matter halo with a constant 
axis ratio.  In this theoretical model, the gas density profiles are 
expressed in terms of the intrinsic properties of the dark matter
halos. Then, we incorporate the projection effect into the gas profiles, 
and show that the gas surface brightness profiles are expressed in terms
of the eccentricities and the orientation angles of the dark halos.  
For the practical purpose of our theoretical model, we 
provide several empirical fitting formulae for the gas density and 
temperature profiles, and also for the  surface brightness profiles 
relevant to X-ray and Sunyaev--Zel'dovich effect observations.  
Finally, we construct a numerical algorithm to determine the halo 
eccentricities and orientation angles using our model, and 
demonstrate that it is possible in principle to reconstruct the 
3D structures of the dark halos from the X-ray and/or Sunyaev-Zel'dovich 
effect cluster data alone without requiring priors such as weak lensing 
informations and without relying on such restrictive assumptions 
as the halo axial symmetry about the line-of-sight.

\end{abstract} 
\keywords{cosmology: theory --- dark matter --- galaxies: clusters: general --- X-rays: galaxies: clusters}

\section{INTRODUCTION}

Understanding of statistical properties of {\it dark matter halos} has
been significantly advanced in recent years, largely owing to the
development of high-resolution numerical simulations 
\citep{nfw96,nfw97,fuk-mak97,moo-etal98,jingsuto00,jingsuto02}.  Given
those theoretical/empirical successes, a next natural question is how to
apply them for the description of {\it real} galaxy clusters.  While
there exist a number of attempts along this line, they usually make 
the unrealistic assumption of the halo spherical symmetry  
\citep{mak-etal98,sut-etal98,yosh-suto99}. This paper describes a
methodology to reconstruct the three-dimensional (3D) structure of dark
halos from the two-dimensional (2D) surface brightness profiles of
intra-cluster gas from the X-ray and/or Sunyaev-Zel'dovich (SZ) effect
observations assuming that the halos are triaxial ellipsoids with a
constant axis ratio.

Indeed it is a classical problem in astrophysics to determine the 3D
properties of astronomical objects from the observed 2D counterparts
\citep[e.g.,][] {lucy74,fabricant84,deh-ger93,bin-etal90,ger-bin96}.
Deprojecting galaxy clusters is one of the most important applications.
\citet{fabricant84} analyzed the X-ray surface brightness map of
clusters, and showed that their mass distribution is far from being 
spherical. \citet{zar-etal98} developed a general method of deprojecting
the 2D images of rich clusters based on the Fourier slice theorem to
reconstruct the 3D cluster structures. Later their technique was 
tested against numerically simulated galaxy clusters
\citep{zar-etal01}.  \citet{yosh-suto99} proposed a deprojection method
for spherical clusters based on the Abel integral.  \citet{reb00}
provided a parameter-free Richardson-Lucy algorithm to reconstruct the
3D halo potential, and demonstrated its stability by applying
it to gas-dynamical simulations.  \citet{dor-etal01}
provided a perturbative approach to the cluster deprojection, taking 
into account the non-isothermality and asphericity of galaxy clusters. 
Recently, \citet{fox-pen02} also considered the problem of deprojecting 
aspherical clusters.  They first constructed a parameterized 3D
axisymmetric cluster model, and determined the 3D cluster shapes 
through the $\chi^{2}$-fitting between the model predictions and the 
simulated cluster data.

All the previous approaches, however, were based on rather restrictive 
assumptions such as the cluster axial symmetry (e.g., oblate or prolate) 
about the line-of-sight, and/or the isothermality of galaxy clusters. 
Furthermore in their approaches, it was required to combine X-ray and/or 
SZ effect data with weak lensing (WL) map, which significantly limits 
the applicability of the previous methods.   In fact, it is generally 
believed that it would be a difficult task to deproject the clusters 
from the 2D projected X-ray or SZ observables alone without having such 
restrictive assumptions and priors given the degeneracy of the
parameters due to the the projection process itself.

Lee \& Suto (2003; hereafter Paper I) suggested that one may understand 
the properties of the dark matter halos from the intra-cluster gas 
profiles. Assuming that the intra-cluster gas is in hydrostatic equilibrium 
in the triaxial dark matter halos,  Paper I derived the 3D density and 
temperature profiles of the intra-cluster gas from the 1st principles 
using the perturbation theory, and found an analytic relation between 
the eccentricities of the intra-cluster gas and the underlying 
dark halo (see eq.[28] in the Paper I).  However, the perturbation
result of the Paper I is valid only in the asymptotic limit of the low
asphericity, and did not include the effect of the projection of the 
gas profiles on the plane of the sky. Here, we generalize the works of 
the Paper I  to the case of highly aspherical clusters, and attempt to
find a general relation between the dark halos and the intra-cluster 
gas taking into account the parameter-degeneracy caused by the
projection. For the practical purpose of our theoretical modeling, 
we provide a series of empirical fitting formulae for the 2D and 3D 
gas profiles which can be used as templates in comparing with the 
observed profiles of clusters, and demonstrate the degree of the
feasibility of the halo reconstruction using the template formulae and 
the simulated surface brightness maps of X-ray and SZ effect.
 
The organization of this paper is as follows.  In \S 2 we construct a 
theoretical model for the intra-cluster gas in hydrostatic equilibrium 
within the gravitational potential of a triaxial dark matter halo with 
a constant axis ratio, and provide a series of empirical fitting formulae 
for the 3D gas profiles in a systematic manner. 
In \S 3 we consider the projection of the 3D gas profiles onto the plane 
of the sky, and provide another series of empirical fitting formulae for 
the 2D surface brightness density profiles.  In \S 4 we describe a 
numerical algorithm to determine the 3D structures of the dark halos 
from the observed 2D cluster surface brightness profiles, and test
the algorithm against a numerical toy model. Finally we discuss the 
final results,  and draw our conclusions in \S 5.

\section{MODELING INTRA-CLUSTER GAS PROFILES}

\subsection{Gravitational Potential of Dark Matter Halos}

To predict theoretically the X-ray and SZ profiles of galaxy clusters,
one first needs a good physical model for the intra-cluster gas. Here we
assume that the intra-cluster gas is either isothermal or polytropic, and
in hydrostatic equilibrium within the gravitational potential generated 
by a concentric and coaxial triaxial dark matter halo.  

Consider a triaxial dark halo whose iso-density surfaces are given by 
the following equation:
\begin{eqnarray}
\label{eq:isodensity}
R^2 \equiv x^2 + \frac{y^2}{1-e_{b}^2} + \frac{z^2}{1-e_{c}^2} ,
\end{eqnarray}
where the Cartesian system of coordinates $(x,y,z)$ is aligned with the
halo principal axes, oriented such that the $x$-axis and $z$-axis 
run along the major and minor principal axes, 
respectively.  The major axis length of the iso-density surface is 
denoted by $R$, while $e_{b}$ and $e_{c}$ ($e_{b} < e_{c}$) represent 
the two constant eccentricities of the ellipsoidal dark halos.  
Equation (\ref{eq:isodensity}) implies that the density profile of a 
triaxial dark matter halo should be a function of the major axis 
length $R$.

We adopt the density profile of a triaxial dark halo proposed by 
\citet{jingsuto02}:
\begin{equation}
\label{eqn:den}
\rho(R) = \frac{\delta_{c}\rho_{\rm crit}}{\left(R/R_0\right)^{\alpha}
\left(1 + R/R_0\right)^{3-\alpha}} ,
\end{equation}
where $R_{0}$ is the scale length, $\delta_{c}$ is the dimensionless
characteristic density contrast with respect to the critical density
$\rho_{\rm crit}$ of the universe at the present epoch, and $\alpha$
represents the inner slope of the density profile.  \citet{jingsuto02}
showed that $\alpha \approx 1$ on the cluster scale and $\alpha \approx
3/2$ on the galaxy scale. For simplicity and definiteness, we focus on
the case of $\alpha = 1$ throughout this paper.

The gravitational potential due to the ellipsoidal halos described by
equation (\ref{eqn:den}) is formally expressed as \citep{bin-tre87}:
\begin{eqnarray}
\label{eqn:phi}
\Phi(\br) &=& -\pi G 
\sqrt{(1-e^{2}_{b})(1-e^{2}_{c})}
\int^{\infty}_{0}\frac{[\psi(\infty)-\psi(m)]}
{\sqrt{(\tau + 1)(\tau + 1 - e^2_{b})(\tau + 1 - e^{2}_{c})}}d\tau, \\
\label{eqn:psi}
\psi(m) &=& 2\int_{0}^{m}\rho(R)RdR, \quad
m^{2} = \frac{x^{2}}{\tau + 1} + 
\frac{y^{2}}{\tau + 1 - e^{2}_{b}} + 
\frac{z^{2}}{ \tau + 1 - e^{2}_{c}}.
\end{eqnarray}
Equations (\ref{eqn:phi}) and (\ref{eqn:psi}) along with equation
(\ref{eqn:den}) allow one to compute the triaxial halo gravitational 
potential at least numerically.  The halo gravitational potential 
$\Phi(\br)$ depends on $\delta_{c}$, $\rho_{\rm crit}$,$R_{0}$ 
as well as  $e_{b}$, $e_{c}$, $\alpha$. However, the dependence of 
$\Phi(\br)$ on $\delta_{c}$, $\rho_{\rm crit}$, and $R_{0}$ can be
separated out by introducing a dimensionless potential $\tPhi$ that 
depends only on $e_{b}$, $e_{c}$ and $\alpha$ such that 
\begin{equation}
\label{eqn:sep}
\tPhi(\br;e_{b},e_{c},\alpha) \equiv 
\frac{\Phi(\br ;e_{b},e_{c},\alpha,\delta_{c},\rho_{\rm crit},R_{0})}
{4\pi G\delta_{c}\rho_{\rm crit}R^{2}_{0}}.
\end{equation}

\subsection{Gas Density and Temperature Profiles in terms of 
Halo Potential}

To determine the intra-cluster gas profiles in terms of the halo potential, 
it is necessary to specify the equation of state for the intra-cluster gas. 
We consider both the isothermal and the polytropic cases in order.

\subsubsection{Isothermal gas} 

The equation of state for the isothermal gas is given as
\begin{equation}
P_{\rm g}(\br) = P_{\rm g0}\frac{\rho_{\rm g}(\br)}{\rho_{\rm g0}} , 
\end{equation}
where $P_{\rm g}$ and $\rho_{\rm g}$ represent the gas pressure and the
gas density, respectively. In what follows, the subscript $0$ of some 
physical variable indicates the value of that physical variable at the 
center $\br = 0$.

The density profile of the isothermal gas in hydrostatic equilibrium
is given as
\begin{equation} 
\label{eqn:iso}
\rho_{\rm g}({\bf r}) = \rho_{\rm g0}
\exp\left[-\kappa\{\tPhi({\bf r})-\tPhi_{0}\}\right] . 
\end{equation} 
We define a dimensionless isothermal gas constant $\kappa$ as 
\begin{equation}
\kappa \equiv 
\frac{4\pi G \mu_{g}m_{p}\delta_{c}\rho_{\rm crit}R^{2}_{0}}
{k_{B}T_{\rm g}} ,
\end{equation}
where $m_{p}$ is the proton mass, $\mu_{g}$ is the mean molecular weight
of the intra-cluster gas, $k_{B}$ is the Boltzmann constant, and $T_{\rm
g}$ is the (constant) gas temperature. 
Introducing $F_{\Phi}(\br)$:
\begin{equation}
\label{eq:fphi_isothermal}
F_{\Phi}(\br) \equiv \exp\left[- \{\tPhi(\br) - \tPhi_{0}\}\right] ,
\end{equation}
one can rewrite equation (\ref{eqn:iso}) as
\begin{equation} 
\label{eq:rhog_isothermal}
\rho_{\rm g}(\br) = \rho_{\rm g0} \, [F_{\Phi}(\br)]^{\kappa} .
\end{equation}

\subsubsection{Polytropic gas} 

The equation of state for the polytropic gas with the polytropic index
of $\gamma$ $(\ne 1)$ is given as 
\begin{equation} 
P_{\rm g}(\br) = P_{\rm g0}\left[\frac{\rho_{\rm g}(\br)}
{\rho_{\rm g0}}\right]^\gamma , 
\end{equation}
and its density and temperature profiles in hydrostatic equilibrium are
found as \footnote{Note that the definition of $\Phi_{0}$ in Paper I is
different from that given here by a constant offset.}
\begin{equation}
\label{eqn:poly}
\rho_{\rm g}(\br) = \rho_{\rm g0}
\left[1 - \kappa_{p}\{\tPhi(\br)-\tPhi_{0}\}\right]^{1/(\gamma-1)}, 
\qquad 
T_{\rm g} = T_{\rm g0} 
\left[1 - \kappa_{p}\{\tPhi(\br)-\tPhi_{0}\}\right] . 
\end{equation}
  
We define a dimensionless polytropic gas constant $\kappa_{p}$ as
\begin{equation}
\kappa_{p} \equiv \frac{\gamma - 1}{\gamma}
\frac{4\pi G\mu_{g}m_{p}\delta_{c}\rho_{\rm crit}R^{2}_{0}}
{k_{B}T_{\rm g0}} , 
\end{equation} 
and introduce $F_{\Phi}(\br)$:
\begin{equation}
\label{eq:fphi_polytropic}
F_{\Phi}(\br) \equiv 1 - [\tPhi(\br) - \tPhi_{0}] .
\end{equation}
Then one can rewrite equation (\ref{eqn:poly}) as
\begin{equation} 
\label{eq:rhog_polytropic}
\rho_{\rm g}(\br) = \rho_{\rm g0} 
\left[1-\kappa_{p} + \kappa_{p}F_{\Phi}(\br) \right]^{1/(\gamma -1)},
\qquad 
T_{\rm g}(\br) = T_{\rm g0} 
\left[1-\kappa_{p} + \kappa_{p}F_{\Phi}(\br) \right] .
\end{equation}

\subsection{Empirical Fitting Formulae for the Gravitational Potential} 

We have shown in $\S 2.2$ that the potential function $F_{\Phi}(\br)$
defines the density and the temperature profiles of intra-cluster gas
completely. In general, $F_{\Phi}(\br)$ does not have a closed
analytic form when the dark matter halos are triaxial ellipsoids.  In
Paper I, we computed $F_{\Phi}(\br)$ analytically with the perturbation
theory assuming $e^{2}_{b} \le e^{2}_{c} \ll 1$, and related the 
iso-density surfaces of the intra-cluster gas to those of the dark
halos; if the halo iso-density surfaces are triaxial ellipsoids with 
the eccentricities of $e_{\sigma}$ ($\sigma=b, c$), 
then the iso-density surfaces of the intra-cluster gas is also well 
approximated as triaxial ellipsoids with the eccentricities 
of $\ep_{\sigma}$ that are related to $e_{\sigma}$ by
\begin{equation}
\label{eqn:ecc}
\frac{\ep^{2}_{\sigma}}{e^{2}_{\sigma}} =
\frac{6(1+u)\ln(1+u) + u^{3}-3u^{2}-6u}{2u^{2}[(1+u)\ln(1+u)-u]},  
\end{equation}
where $u \equiv \vert \br \vert/R_{0}$. Note that the right-hand-side of
equation (\ref{eqn:ecc}) are written in terms of the rescaled 
spherical radius $u$ instead of the rescaled major axis length, 
which results from the fact that we neglected the higher-order
terms $O(e_\sigma^4)$ in deriving equation (\ref{eqn:ecc}) with  
the perturbation theory.  Although equation (\ref{eqn:ecc}) seems
complicated, it is a slowly varying function, and well approximated as a
constant in the range of $0< u <1$ (see Fig. 3 of Paper I) such that
\begin{equation} 
\label{eqn:newecc}
\frac{\ep^{2}_{\sigma}}{e^{2}_{\sigma}} \approx 0.7^{2} .   
\end{equation}
 
Strictly speaking, equation (\ref{eqn:newecc}) is valid only in the
limit of $e^{2}_{b} \le e^{2}_{c} \ll 1$ over the range of $0< u <1$.
To proceed further in a tractable fashion, however, we extend the validity 
of equation (\ref{eqn:newecc}) to $e^{2}_{b} \le e^{2}_{c} \le 1$ over 
the whole range of $u$.  In other words, we assume that the iso-density 
surfaces of the intra-cluster gas are triaxial ellipsoids with the constant 
eccentricities of $\ep_{b},\ep_{c}$ that are related to the halo 
eccentricities $e_{b},e_{c}$ by equation (\ref{eqn:newecc}). Actually 
this turns out to be a good approximation as will be shown below. 

This assumption allows us to write the potential function
$F_{\Phi}(\br)$ in terms of the major axis length of the gas iso-density
surfaces, $\xi$, defined as
\begin{equation}
\label{eqn:isogas}
\xi^{2} \equiv u^2_{x} + \frac{u^2_{y}}{1-\ep_b^2} 
+ \frac{u^2_{z}}{1-\ep_c^2} 
= \frac{1}{R_0^2} \left(  x^2 + \frac{y^2}{1-\ep_b^2} 
+ \frac{z^2}{1-\ep_c^2} \right) . 
\end{equation}
After some trials and errors, we find that the following empirical
formula fits both equations (\ref{eq:fphi_isothermal}) and
(\ref{eq:fphi_polytropic}) quite well:
\begin{equation}
\label{eqn:3dmodel}
F_{\Phi}(\xi) = \left(\frac{1 + \eta \xi^{p}}{1 + \beta
\xi^{p}}\right)^{q},   
\end{equation}
where $\beta$, $p$, $q$, and $\eta$ are free parameters.  The four free 
parameters are not constant but supposed to be functions of the halo 
eccentricities.  We determine the functional form of each parameter 
by taking the following steps.
\begin{enumerate}
\item Compute $F_{\Phi}(\br)$ numerically using equations
      (\ref{eqn:phi}) and (\ref{eqn:psi}) for various values of $e_{b}$
      and $e_{c}$.
\item Fit those numerical data points to the empirical model
      (eq. [\ref{eqn:3dmodel}]) and obtain the best-fit values of
      $\beta$, $p$, $q$, and $\eta$ using the Levenberg-Marquardt
      method \citep{pre-etal92}. 
\item Finally model those best-fit parameters as functions of $e_{b}$
      and $e_{c}$. In practice, we find that all the parameters can 
      be written as functions of a single variable 
      $\mu \equiv e^{3}_{b} + e^{3}_{c}$.
\end{enumerate} 

\begin{figure}
\begin{center}
\epsscale{1.0} 
\plotone{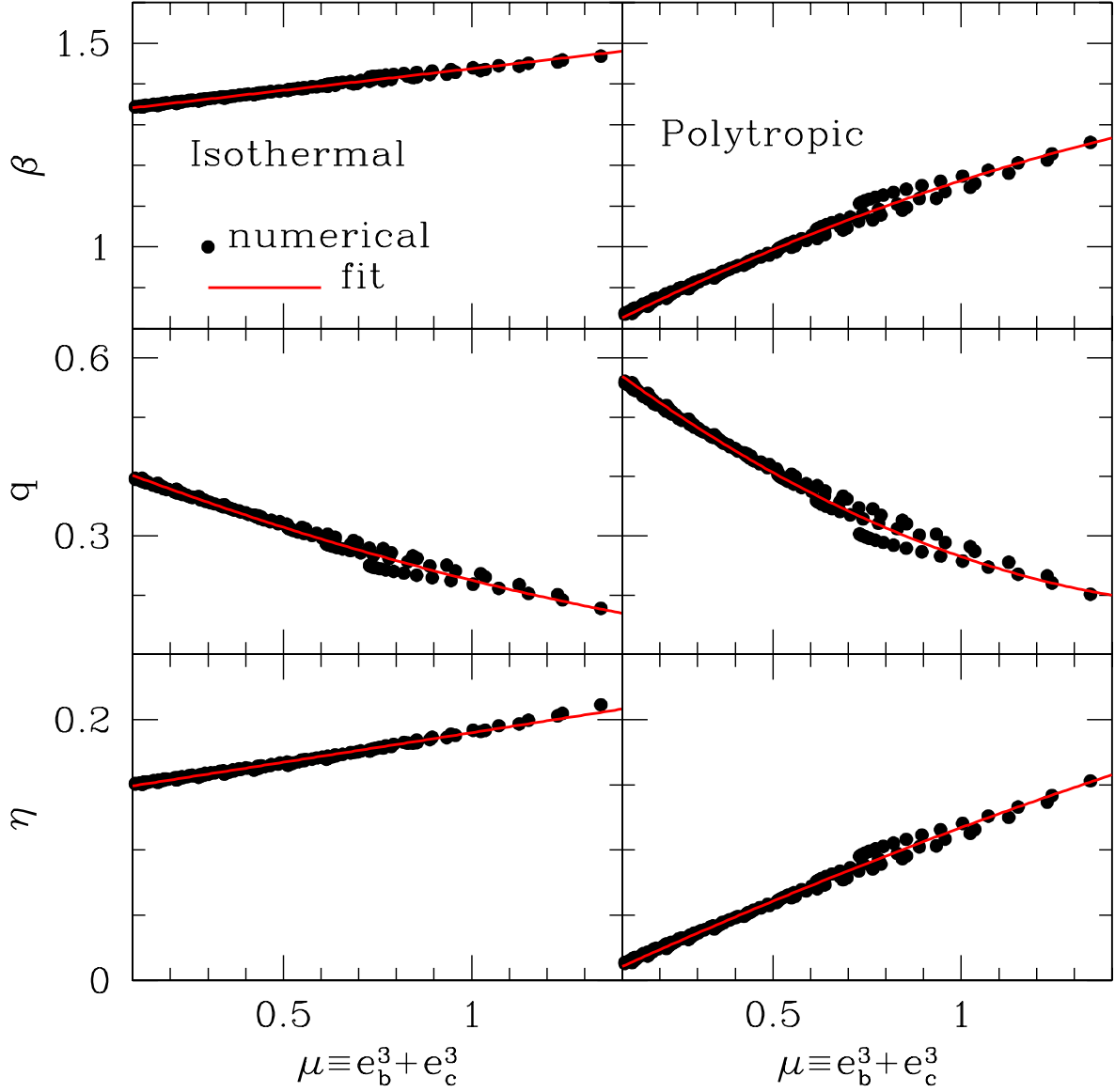}
\caption{Parameters describing our empirical model of the halo
potential profile (eq.[\ref{eqn:3dmodel}]) as a function of the halo
eccentricities ($\mu \equiv e^{3}_{b} + e^{3}_{c}$).  The filled circles
represent the best-fit values to the numerical results at each $\mu$,
and the solid curves show the corresponding polynomial fitting curves.
The intra-cluster gas is assumed to be isothermal ({\it Left}) and
polytropic ({\it Right}).
\label{fig:3dparameter}}
\end{center}
\end{figure}

\begin{figure}
\begin{center}
\plotone{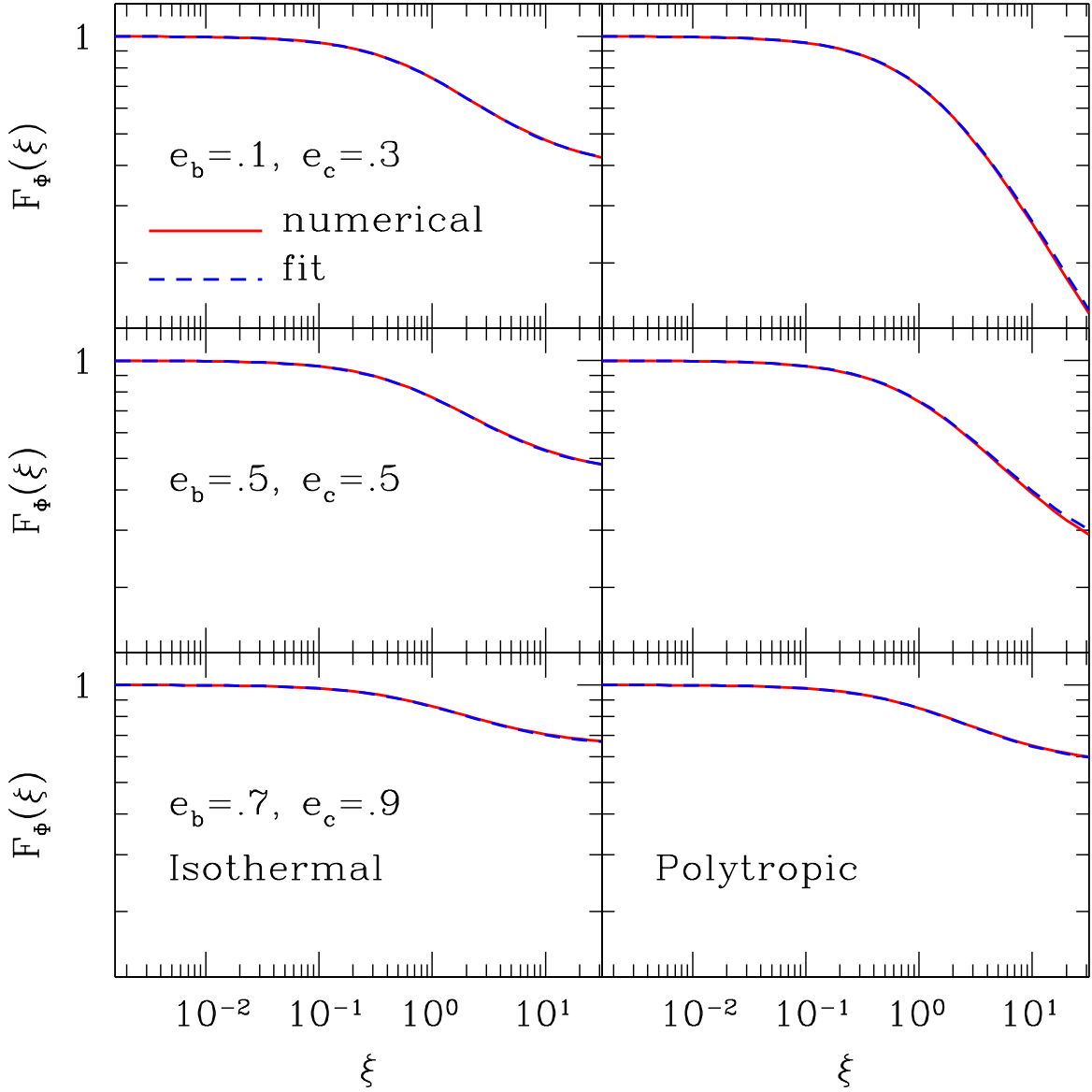}
\caption{Profiles of gravitational potentials  of dark halos.
Solid and dashed lines indicate the results of the direct numerical
integration and our empirical fitting model, respectively.  The
intra-cluster gas is assumed to be isothermal ({\it Left}) and
polytropic ({\it Right}).  The eccentricities of the underlying dark
halos are {\it Top}: $e_{b}=0.1$ and $e_{c}=0.3$, {\it Middle}:
$e_{b}=0.1$ and $e_{c}=0.7$, {\it Bottom}: $e_{b}=0.7$ and $e_{c}=0.9$.
\label{fig:3dprofile}}
\end{center}
\end{figure}

We find the following polynomials to provide good fits: 
\begin{eqnarray}
\label{eqn:beta}
\beta &=& c_{\beta 0} + c_{\beta 1}\mu, \\
\label{eqn:p}
p     &=& const. , \\ 
\label{eqn:q}
q     &=& c_{q 0} + c_{q 1}\mu + c_{q 2}\mu^{2}, \\ 
\label{eqn:eta} 
\eta  &=& c_{\eta 0} + c_{ \eta 1}\mu .
\end{eqnarray}
Table \ref{table:3d_coef} lists the best-fit values of the polynomial
coefficients. The best-fit values of the constant $p$ is determined to
be unity for all cases. 
Figure \ref{fig:3dparameter} plots $\beta$, $q$, and $\eta$ as a
function of $\mu$.  The filled circles indicate the best-fit values of
the parameters, while the solid lines represent their polynomial fits.
Note that $\eta$ is one order-of-magnitude smaller than $\beta$ for all
cases, suggesting that the term associated with $\eta$ can be safely
neglected except at the outer part of clusters.  The accuracy of the
fits is illustrated in Figure \ref{fig:3dprofile} where $F_{\Phi}(\xi)$
is plotted against $\xi$. The solid lines represent the numerical
results while the dashed lines correspond to equation
(\ref{eqn:3dmodel}) with the best-fit coefficients of the polynomials
plotted in Fig. \ref{fig:3dparameter} and listed in Table
\ref{table:3d_coef}.  Our empirical fitting formulae reproduce the
numerical results very well for all cases over a wide range of $\xi$
within a fractional error less than $0.5\%$.  

The characteristic feature of our fitting model is that all the
parameters are expressed in terms of $\mu=e^{3}_{b}+e^{3}_{c}$.  
Given that the first-order perturbation approach of the Paper I 
indicated the dependence of the halo potential on $e^{2}_{b}+e^{2}_{c}$, 
one may have expected  the fitting model to depend on
$e^{2}_{b}+e^{2}_{c}$ rather than $e^{3}_{b}+e^{3}_{c}$.
As a matter of fact,  we indeed attempted first to model the parameters
in terms of $e^{2}_{b}+e^{2}_{c}$. But, it turned out that the
parameter values do not scale in terms of $e^{2}_{b}+e^{2}_{c}$ 
while they show very good scaling feature in terms of $e^{3}_{b}+e^{3}_{c}$ 
as shown in Figure \ref{fig:3dparameter}.
We have not yet completely  understood the origin of this scaling, 
but the result is empirically quite robust. 

In addition, we note that in the Paper I we derived the gas-halo 
eccentricity relation (eq. [\ref{eqn:ecc}]) by using the first-order 
perturbation theory, and showed that the approximation error can be 
expressed as a function  of $e^{3}_{b}$ and $e^{3}_{c}$ 
(see eqs [29] and [30] in the Paper I). 
The dependence of the approximation error on $e^{3}_{b}$ and $e^{3}_{c}$ 
implies that the higher order terms neglected in the first-order
perturbation result may be scaled in terms of $e^{3}_{b}$ and
$e^{3}_{c}$.  Now that we have used  equation (\ref{eqn:ecc}) to
construct the fitting model (eq. [\ref{eqn:3dmodel}]),  one may expect
that equation (\ref{eqn:3dmodel}) has similar scaling.  In other words, 
we suspect that the dependence of equation (\ref{eqn:3dmodel}) on 
$e^{3}_{b}+e^{3}_{c}$ rather than $e^{2}_{b}+e^{2}_{c}$ may be related 
to the scaling of the higher order terms neglected in the first-order 
perturbation result of the Paper I. 

Note also that in our model the gas density/temperature profiles do 
not approach zero even at large $u$.  This feature is ascribed to the 
hydrostatic equilibrium condition itself.  In reality, the hydrostatic 
equilibrium condition may be satisfied only within some radius, e.g., 
the virial radius of the cluster. In fact, the halo density profiles 
at those regions are not well approximated by equation (\ref{eqn:den}) 
because of the presence of substructures, and so on.  
This problem, however, is unlikely to limit the validity of our 
methodology in practice since the X-ray and SZ fluxes from regions 
beyond the cluster virial radius are usually negligible.

\section{PROJECTED SURFACE BRIGHTNESS PROFILES OF INTRA-CLUSTER GAS}

Now we are in a position to model the X-ray and SZ surface brightness 
profiles of triaxial galaxy clusters by incorporating the projection 
effect into the 3D models developed in $\S 2$. 
Just for comparison, let us recall that the surface brightness 
profiles  of spherical clusters can be readily evaluated as  
\begin{equation}
\Sigma (\theta) = 
\int_{-\infty}^\infty L(\sqrt{d_{\scriptscriptstyle A}^2\theta^2+z^2}) dz , 
\end{equation}
where $\theta$ is the angular radius from the center of the cluster,
$d_{\scriptscriptstyle A}$ is the angular diameter distance to the
cluster, and $L$ is the emissivity given in terms of $\rho_{g}$ and
$T_{g}$ such that $L \propto \rho^{2}_{\rm g}T^{1/2}_{\rm g}$ for
bolometric X-ray and $L \propto \rho_{\rm g}T_{\rm g}$ for SZ
observations, respectively.  For the case of a triaxial cluster, however,
the result becomes much more complicated because the 2D projection 
depends on the relative direction of the line-of-sight of an observer
with respect to the principal axes of a triaxial halo.  
The projection effect of triaxial bodies on the plane of the sky is 
fully discussed by \citet{stark77} and \citet{binney85}.  
In what follows we adopt the notation of \citet{binney85}.
 
Let $(\theta,\phi)$ be the polar angle of the line-of-sight in the halo 
principal coordinate system $(x,y,z)$, and let the observer's coordinate 
system be defined by Cartesian axes $(\xp,\yp,\zp)$ with $\zp$-axis aligned 
with the line-of-sight direction and $\xp$-axis lying in the $(x,y)$ 
plane \citep[see Fig. 1 in][]{oguri-etal03}.
Then the halo principal coordinate system $(x,y,z)$ is related to the
observer's coordinate system $(\xp,\yp,\zp)$ by
\begin{eqnarray}
x &=& -\sin\phi\xp - \cos\phi\cos\theta\yp + \cos\phi\sin\theta\zp, \\
y &=& \cos\phi\xp - \sin\phi\cos\theta\yp + \sin\phi\sin\theta\zp, \\
z &=& \sin\theta\yp + \cos\theta\zp.
\end{eqnarray}

If we use the major axis length $\xi$ defined in equation
(\ref{eqn:isogas}) to characterize the iso-density surfaces, then the
projection of $L$ onto the plane of the sky is written as
\begin{equation}
\label{eqn:proj}
\Sigma(\xp,\yp) \equiv \int^{\infty}_{-\infty}L(\xi^2)d\zp = 
\frac{2}{\sqrt{f}}\int^{\infty}_{0}L(z^{\prime\prime 2}  +
\lam^{2})
dz^{\prime\prime }, 
\end{equation}
where 
\begin{eqnarray}
\label{eqn:zpp}
z^{\prime\prime} &=& \sqrt{f}\left(\zp + \frac{g}{2f}\right), \\
\label{eqn:lam}
\lam &=& \frac{1}{\sqrt{f}}(Ax^{\prime 2} + 
Bx^{\prime }y^{\prime } + Cy^{\prime 2})^{1/2}, \\
\label{eqn:f}
f &=& \sin^{2}\theta\left(\cos^{2}\phi + 
\frac{\sin^{2}\phi}{1-\ep_{b}^{2}}\right) 
+ \frac{\cos^{2}\theta}{1-\ep_{c}^{2}}, \\
\label{eqn:g}
g &=& \sin\theta\sin2\phi\left(\frac{1}{1-\ep_{b}^{2}}-1\right)\xp + 
\sin 2\theta\left(\frac{1}{1-\ep_{c}^{2}}-\cos^{2}\phi-
\frac{\sin^{2}\phi}{1-\ep_{b}^{2}}\right)\yp, \\
\label{eqn:A}
A &=& \frac{\cos^{2}\theta}{1-\ep_{c}^{2}}
\left(\sin^{2}\phi + \frac{\cos^{2}\phi}{1-\ep_{b}^{2}}\right)
+ \frac{\sin^{2}\theta}{1-\ep_{b}^{2}}, \\
\label{eqn:B}
B &=& \cos\theta\sin2\phi\left(1 - \frac{1}{1-\ep_{b}^{2}}\right)
\frac{1}{1-\ep_{c}^{2}}, \\
\label{eqn:C}
C &=& \left(\frac{\sin^{2}\phi}{1-\ep_{b}^{2}} + \cos^{2}\phi\right)
\frac{1}{1-\ep_{c}^{2}}. 
\end{eqnarray}
Hence $\Sigma(\xp,\yp) =\Sigma(\lam)$. In other words, the 2D isophotal
curves of triaxial galaxy clusters are concentric and coaxial ellipses
if their eccentricities $\ep_{\sigma}$ of the {\it gas} iso-density surfaces 
are constants. As we noted before, however, it is not exactly the case 
for our model where the eccentricities $e_{\sigma}$ of the {\it halo}
iso-density surfaces are constant. Nevertheless this holds approximately
(see eq.[\ref{eqn:newecc}]).

Let us define a dimensionless surface brightness $\tSigma \equiv
\Sigma/\Sigma_{0}$ where $\Sigma_{0} \equiv \Sigma(0)$ and $\tlam =
\lam/R_{0}$.  In the same spirit as for the 3D model (\ref{eqn:3dmodel}), 
we propose the following empirical model for the 2D surface brightness 
profiles:
\begin{equation}
\label{eqn:2dmodel}
\tSigma(\tlam;e_{b},e_{c},\kappa,\kappa_{p},\gamma) = \left(\frac{1 +
\omega\tlam^{s}}{1 + \zeta\tlam^{s}}\right)^{t}.  
\end{equation}
Note that $\tSigma$ depends not only on $\mu$ but also on $\kappa$ 
(or $\kappa_{p}$), and even on the polytropic index $\gamma$ for 
the polytropic case. Thus, we consider the isothermal and polytropic 
cases separately, and obtain the following empirical fits. 

For the isothermal gas, the parameters are fitted to the
following polynomials of $\mu$ and $\kappa$: 
\begin{eqnarray}
\label{eq:2dparam_iso}
\omega &=& \sum_{i,j=0}^2 \omega_{ij} \mu^i \kappa^j, \\
\zeta &=& \sum_{i,j=0}^2 \zeta_{ij} \mu^i \kappa^j, \\
s     &=& const. , \\ 
t &=& \sum_{i,j=0}^2 t_{ij} \mu^i \kappa^j . 
\end{eqnarray}
For the polytropic gas, we fix $\gamma$ to $\gamma = 1.15$ following
\citet{kom-sel01}, and fit the parameters to the following polynomials  
of $\mu$ and $\kappa$: 
\begin{eqnarray}
\label{eq:2dparam_poly}
\omega &=& \sum_{i,j=0}^3 \omega_{ij} \mu^i \kappa_p^j, \\
\zeta &=& \sum_{i,j=0}^3 \zeta_{ij} \mu^i \kappa_p^j, \\
s     &=& const. , \\ 
t &=& \sum_{i,j=0}^3 t_{ij} \mu^i \kappa_p^j . 
\end{eqnarray}
 
We determine the best-fit values of $\omega_{ij}$, $\zeta_{ij}$, and
$t_{ij}$ using the same method as in $\S 2$. First we calculate $\tSigma$ 
of equation (\ref{eqn:proj}) numerically, approximating the integration 
$\int^{\infty}_{-\infty}L d\zp$ to $\int^{z_{c}}_{-z_{c}}L d\zp$  
where $z_{c} = 20R_{0}$ which is roughly twice the virial radius of 
galaxy clusters \citep{mak-etal98}.  Then we compare the numerical data 
points of $\tSigma$ with the fitting formula (\ref{eqn:2dmodel}), 
and determine the best-fit values for each point using the 
Levenberg-Marquardt method.  Finally we model the free parameters as 
polynomials of $\mu$ and $\kappa$ (or $\kappa_{p}$), and determine 
the best-fit polynomial coefficients.

\begin{figure}
\begin{center}
\plotone{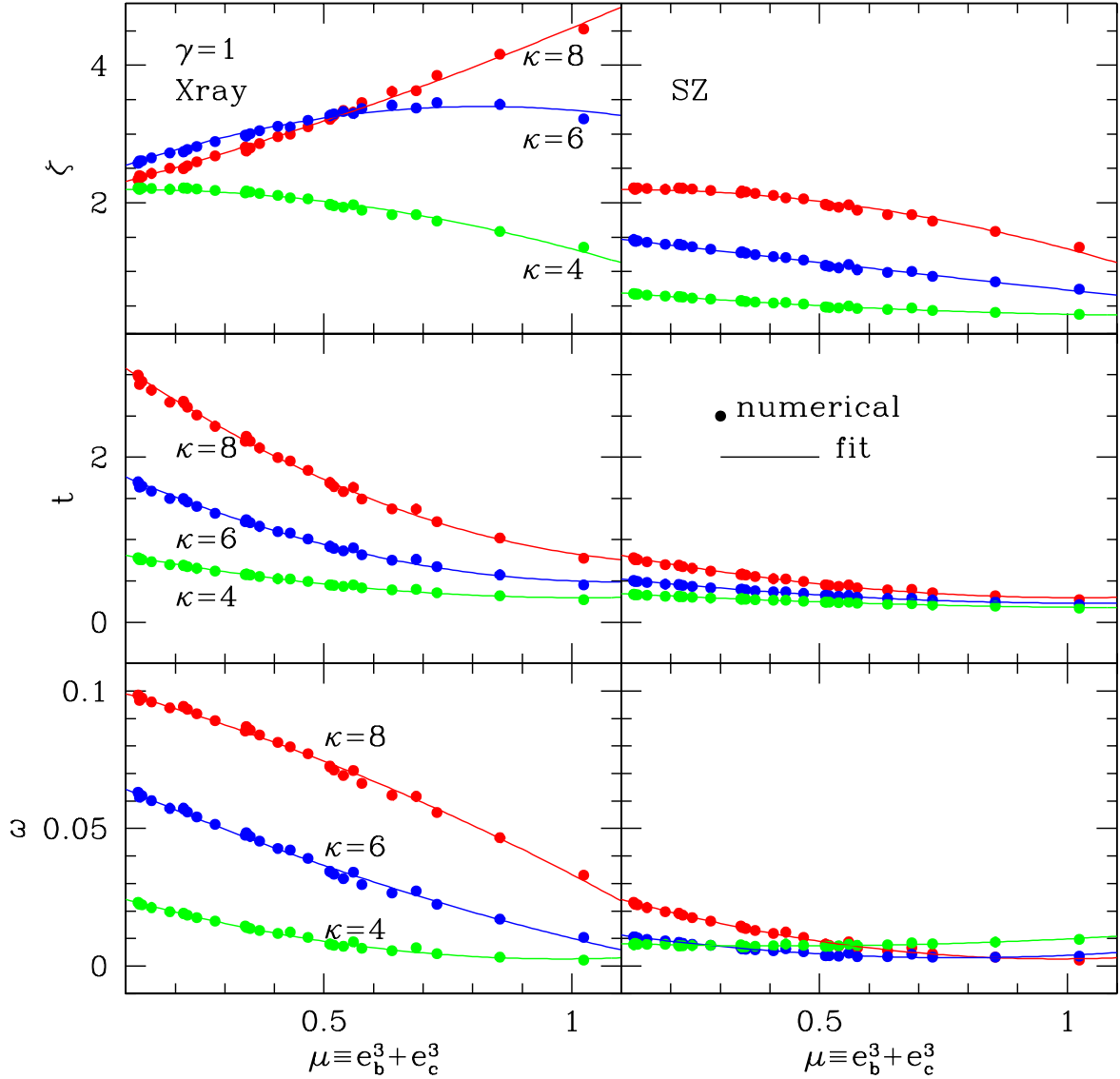} 
\caption{Parameters describing our empirical model of
the surface brightness profile (eq.[\ref{eqn:2dmodel}]) as a function of
$\mu$ for the isothermal case. 
The filled circles
represent the best-fit values to the numerical results at each $\mu$,
and the solid curves show the corresponding polynomial fitting curves.
{\it Left}:  X-ray. 
{\it Right}: SZ.
\label{fig:2disopara}}
\end{center}
\end{figure}

\begin{figure}
\begin{center}
\plotone{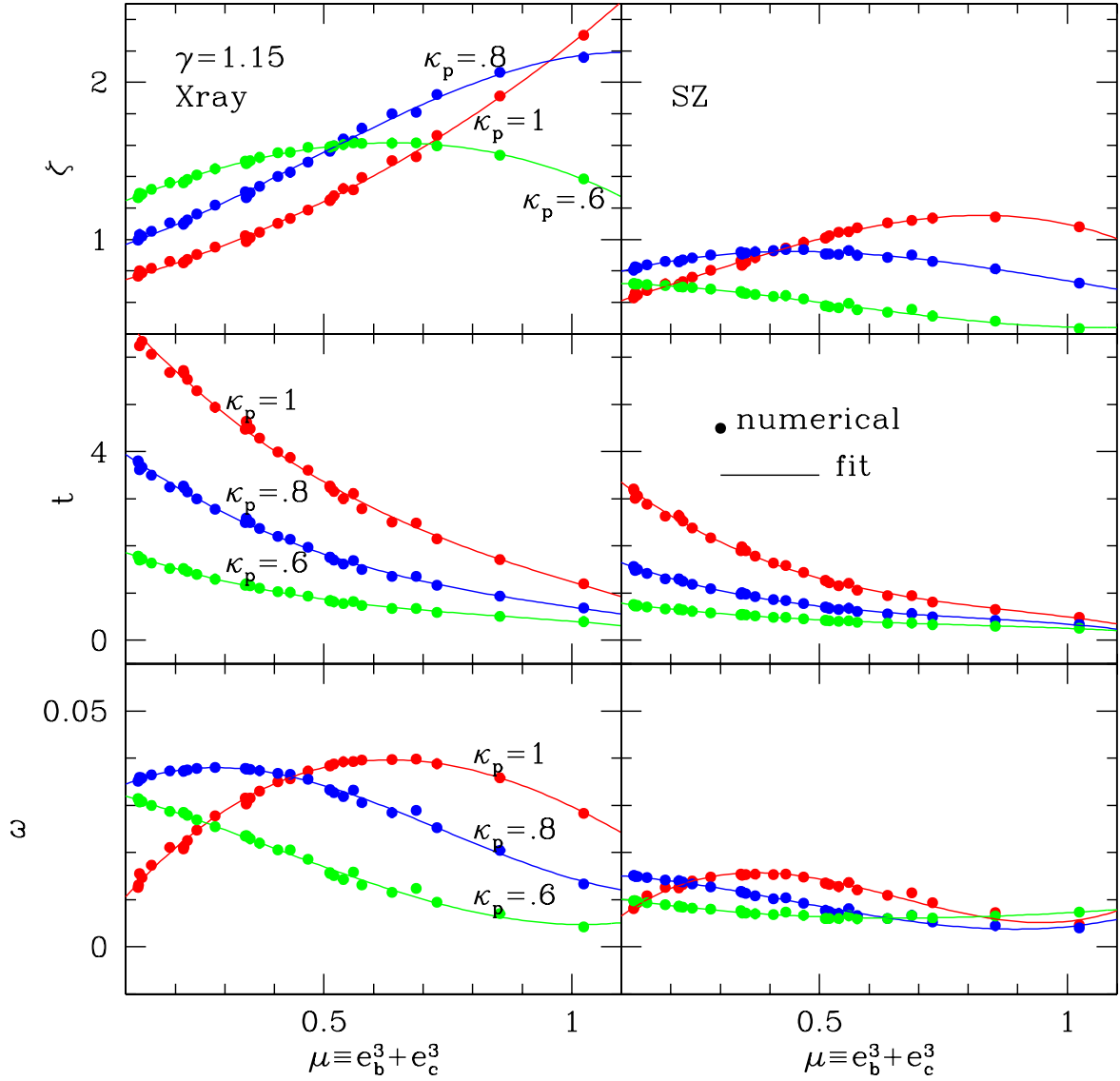}
\caption{Same as Figure \ref{fig:2disopara} but for the 
polytropic case. 
\label{fig:2dpolypara}}
\end{center}
\end{figure}

The best-fit constant values of $s$ is determined to be unity for both
the isothermal and polytropic cases just like the 3D model.  Figures
\ref{fig:2disopara} and \ref{fig:2dpolypara} plot the best-fit
parameters as functions of $\mu$ for three different values of $\kappa$
(or $\kappa_p$). It is clear from these figures that the polynomial
fitting works quite well for all cases.
Tables \ref{table:iso_coef} and \ref{table:poly_coef} list the best-fit
coefficients $\zeta_{ij},t_{ij},\omega_{ij}$ for the cases of bolometric
X-ray and SZ observations, respectively.

\begin{figure}
\begin{center}
\plotone{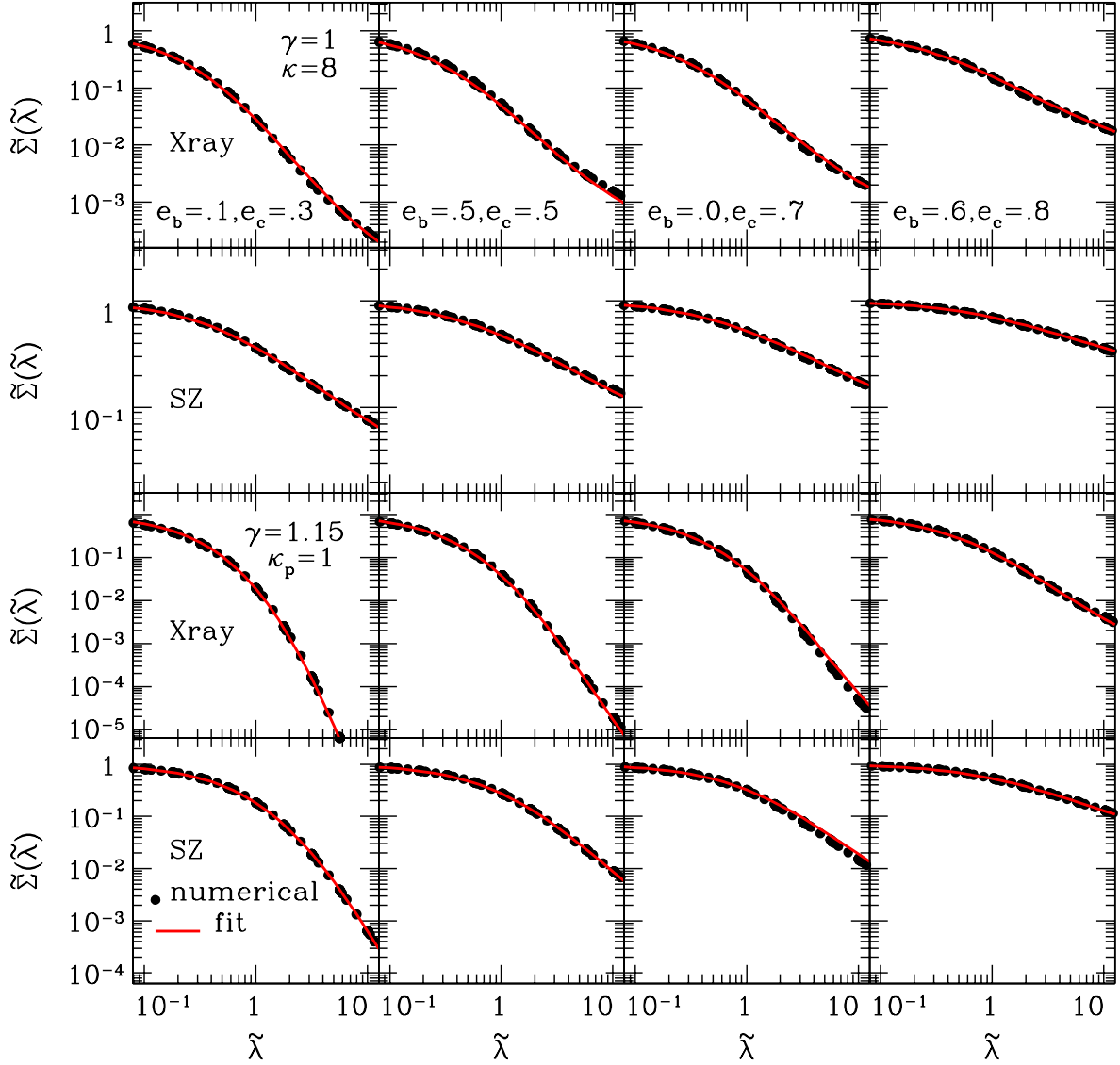} 
\caption{Profiles of the cluster surface brightness. The filled circles
are computed numerically, while our fitting models are shown in solid
lines. From left to right panels, we adopt the underlying halo
eccentricities of $(e_{b}, e_{c})=$ (0.1, 0.3), (0.5,0.5), (0.0, 0.7)
and (0.6, 0.8).  From top to bottom panels, we show the cases of
isothermal X-ray ($\kappa = 8$), isothermal SZ ($\kappa = 8$),
polytropic X-ray ($\gamma = 1.15$ and $\kappa_{p}=1$), and polytropic SZ
($\gamma = 1.15$ and $\kappa_{p}=1$) observations.
\label{fig:2dprofile}}
\end{center}
\end{figure}

Figure \ref{fig:2dprofile} illustrates the degree of the accuracy of our
fitting formulae.  The filled squares represent the numerical results
while the solid lines represent our fitting models with the
best-fit polynomials of $\zeta$, $t$, and $\omega$.  We have found that
our fits reproduce the numerical results within the fractional error of
$20\%$ for all cases in a range of $0 \le \tlam \le 10$.  

\section{DARK HALO RECONSTRUCTION}

In the triaxial dark halo model that we adopt here, the halo
reconstruction from the X-ray and SZ cluster observations is basically to
determine the shapes ($e_{b},e_{c}$) and the orientations ($\theta,\phi$) 
of the dark halos from the observed surface brightness profiles.

In \S 3, we construct a parameterized model for the X-ray and SZ
surface brightness profiles of galaxy clusters. The model is expressed
as a function of the rescaled major axis length of the gas isophotes and
characterized by three parameters.  It turns out that the rescaled major
axis length and the three parameters depend on the eccentricities and
the orientation angles of the underlying dark halos, as well as the gas
constant.  Therefore, we find a way to link the observed 2D surface
brightness profiles of galaxy clusters to the 3D structures of the 
underlying dark halo and the gas constants of the intra-cluster gas 
as well.  In other words, one may expect to determine the values of 
$(e_{b},e_{c}, \theta,\phi)$ and $(\kappa,\kappa_{p})$ by fitting our 
2D parameterized models (eqs. [\ref{eqn:2dmodel}] --
[\ref{eq:2dparam_poly}]) to the X-ray and SZ cluster data.

To demonstrate how our modeling of cluster profiles can be used in the
determination of the shapes and the orientations of dark matter halos,
we apply the following reconstruction algorithm to the numerically
projected profiles.  We first numerically compute $\tSigma$ directly
using equations (\ref{eqn:phi})-(\ref{eqn:proj}).  Then we construct a
pixeled map of surface brightness in $N\times N$ grids ($N=32$ in the
present example) corresponding to the linear scale of $-10R_{0} \le
\xp,\yp\le 10R_{0}$ since $10R_{0}$ is roughly equal to the cluster
virial radius \citep{jingsuto02}.  We create several realizations of
X-ray and SZ profiles for both isothermal and polytropic cases using 
various different values of $e_{b},e_{c},\theta,\phi$.

Our reconstruction algorithm proceeds as follows;
\begin{itemize}
\item 
At each pixel point, say, $(\xp_{i},\yp_{j})$, build the model 
$\tSigma^{\rm th}(\xp_{i},\yp_{j})$ using equation (\ref{eqn:2dmodel}). 
The model is characterized by five free parameters 
$e_{b},e_{c}, \theta,\phi,\kappa$ (or $\kappa_p$) at each point. 
\item
Fit $\tSigma^{\rm th}$ to the observed surface brightness density profiles 
$\tSigma^{obs}$, and calculate the $\chi^{2}$: 
\begin{equation}
\chi^{2} = \sum_{i,j=1}^{N} 
\frac{[\tSigma^{\rm th}(\xp_{i},\yp_{j})-\tSigma^{obs}(\xp_{i},\yp_{j})]^{2}}
{\sigma^{2}_{ij}},
\end{equation}
where $\sigma_{ij}$ denotes the observational error at each pixel
point $(\xp_{i},\yp_{j})$. Those systematic/random errors should
depend on specific observation conditions and thus are not easy to
estimate a priori. Thus in the following numerical tests, we simply
assume that $\sigma_{ij}$ is unity, independent of  $i$ and $j$. 
\item 
Determine the best-fit values of $e_{b},e_{c}, \theta,\phi,\kappa$ (or
$\kappa_p$) through the $\chi^{2}$ minimization.
\end{itemize}

\begin{figure}
\begin{center}
\plotone{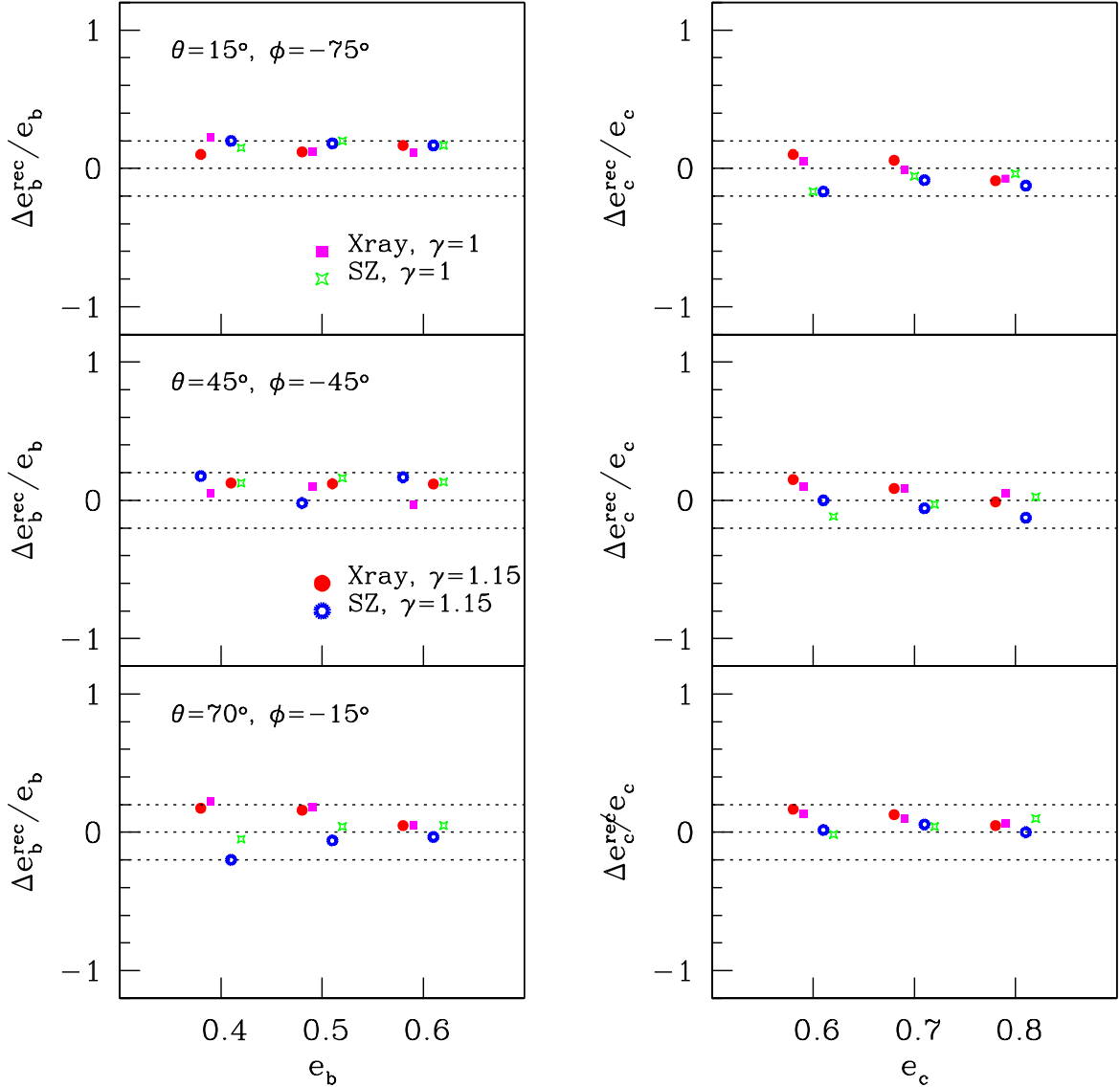} 
\caption{Fractional errors of the reconstructed halo eccentricities.
Filled squares, open squares, filled circles, and open circles
correspond to the cases of isothermal X-ray, isothermal SZ, polytropic
X-ray, and polytropic SZ, respectively.  We adopt the orientation angles
of the line-of-sight with respect to the cluster in the halo principal
frame as $(\theta, \phi) =$ $(15^{o}, -75^{o})$, $(45^{o}, -45^{o})$,
and $(75^{o}, -15^{o})$ from top to bottom panels.
\label{fig:recon_ebc}}
\end{center}
\end{figure}

\begin{figure}
\begin{center}
\plotone{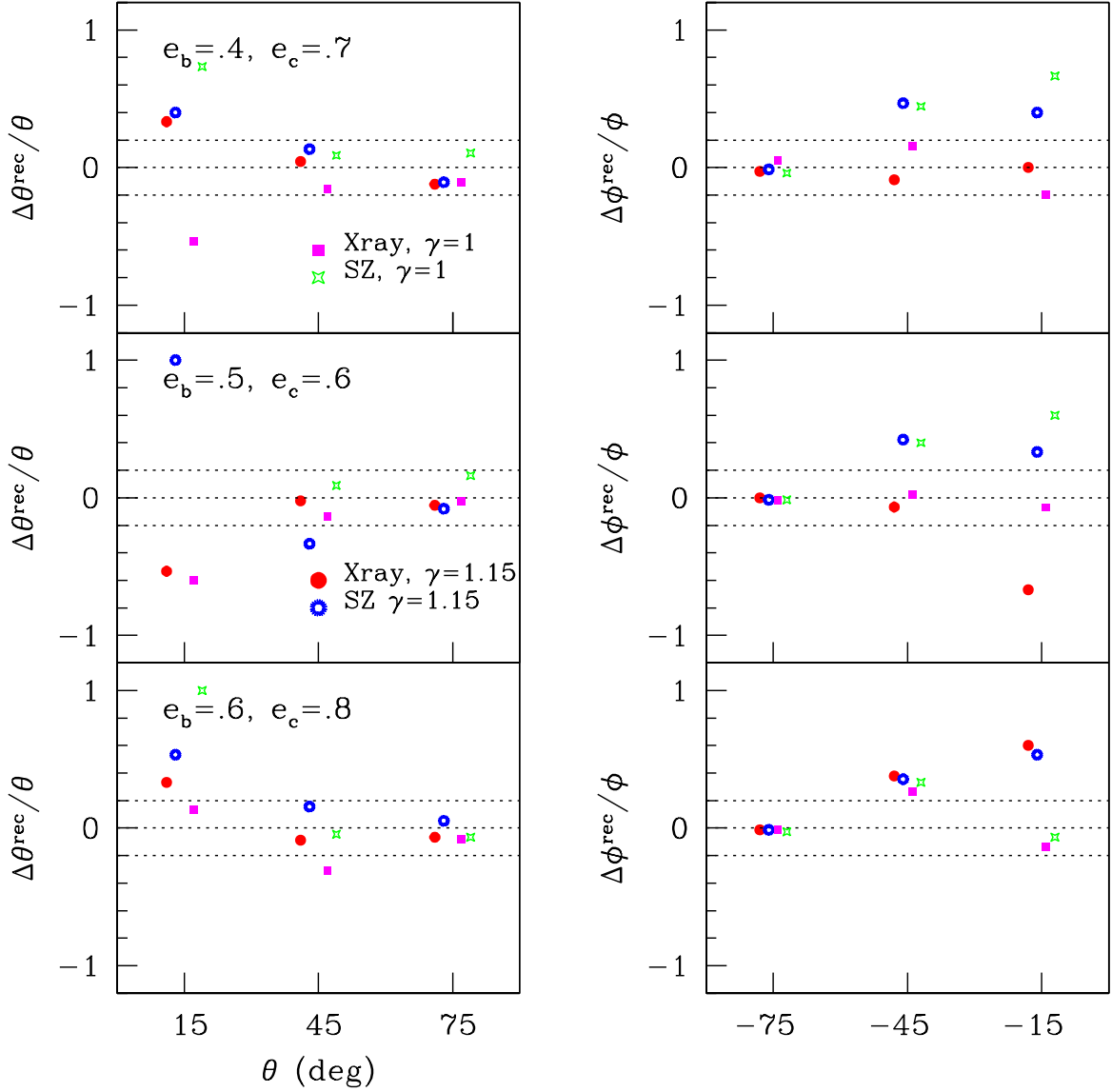} 
\caption{Fractional errors of the reconstructed orientation angles.
Filled squares, open squares, filled circles, and open circles
correspond to the cases of isothermal X-ray, isothermal SZ, polytropic
X-ray, and polytropic SZ, respectively.  We adopt the halo eccentricities
of $(e_{b}, e_{c})=$ = (0.4, 0.7), (0.5, 0.6), and (0.6, 0.8)
from top to bottom panels.
\label{fig:recon_tp}}
\end{center}
\end{figure}

The numerical testing results have revealed that the above algorithm works 
quite well in reconstructing the halo eccentricities within the
fractional error of $20 \%$ as long as the halo eccentricities are not 
so small ($e_{b},e_{c} > 0.3$), which can be understood considering that
our reconstruction algorithm strongly relies on the non-spherical 
signature, and thus is expected to fail for the case of almost spherical 
halos with low eccentricities ($e_{b},e_{c} < 0.3$).  For the low
eccentricity cases, the perturbation results of the Paper I may be 
more useful. Since the low-eccentricity case is not the main interest 
of the present paper, we do not investigate those cases here. 
The algorithm also works in reconstructing the orientation angles 
but suffer relatively large fractional errors.  We show
the examples of our numerical reconstruction results in Figures
\ref{fig:recon_ebc} and \ref{fig:recon_tp} for the following cases of
the halo eccentricities and orientation angles: $(e_{b},e_{c}) =
(0.4,0.7)$, $(0.5,0.6)$, and $(0.6,0.8)$; $(\theta,\phi) =
(15,-75)$,$(45,-45)$, $(75,-15)$ in units of degree.
 
Figure \ref{fig:recon_ebc} plots the fractional error of the reconstructed
eccentricities versus their input values for three different cases of
the orientation angles.  The left three panels are for $e_{b}$, and the
right three panels for $e_{c}$. The top two panels correspond to the
case of $\theta = 15^{o}$ and $\phi = -75^{o}$, the middle two panels to
the case of $\theta = 45^{o}$ and $\phi = -45^{o}$, and the bottom three
panels to the case of $\theta = 75^{o}$ and $\phi = -15^{o}$.  For
almost all cases, the fractional errors fall within $20 \%$.

Figure \ref{fig:recon_tp} plots the fractional error of the reconstructed 
orientation angles versus the original orientation angles for three 
different cases of the eccentricities. The left three panels are for 
$\theta$, and the right three panels for $\phi$. The top two panels 
correspond to the case of $e_{b} = 0.4$ and $e_{c} = 0.7$, the middle 
two panels to the case of $e_{b} = 0.5$ and $e_{c} = 0.6$, and the bottom 
three panels to the case of $e_{b} = 0.6$ and $e_{c} = 0.8$.  For most cases, 
the fractional errors fall within $20 \%$. However, for a few cases, 
the fractional errors are larger than $50 \%$, indicating that the 
angle reconstruction is not always stable compared with the eccentricity 
reconstruction.

\section{DISCUSSIONS AND CONCLUSIONS}

It was regarded as a difficult task to reconstruct the 3D structures 
of dark matter halos from the projected 2D surface-brightness maps 
of X-ray and/or SZ clusters given the parameter-degeneracy caused by 
the projection process itself. To break the degeneracy, previous 
approaches had to rely on such restrictive assumptions as 
the cluster axial symmetry about the line-of-sight and some priors 
like weak lensing informations. Here we have developed, for the first
time, a theoretical framework within which the 3D halo reconstruction 
is possible in principle without such restrictive assumptions and 
additional priors. 

We derived the density profiles of the intra-cluster gas from the 
1st principles, assuming that the intra-cluster gas is either isothermal 
or polytropic, and in hydrostatic equilibrium within the gravitational 
potential of concentric and coaxial triaxial dark matter halos. 
In our theoretical modeling of the intra-cluster gas, the density 
profiles depend explicitly on the intrinsic properties of the 
underlying halos (the two halo eccentricities, $e_{b}$ and $e_{c}$).
For the case of highly aspherical halos, however, we have found that 
the gas density profiles cannot be expressed in closed analytic 
forms.  Therefore, we have attempted to find general fitting formulae 
for the gas density profiles as functions of the halo eccentricities 
which may be applicable even to highly aspherical clusters.  
We have found empirically a simple scaling relation with respect to 
$\mu = e^{3}_{b} + e^{3}_{c}$, and provided a set of fitting 
formulae for the 3D gas density profiles expressed in terms 
of $\mu$. 

Then, we have incorporated the projection effect into the gas 
density profiles, and provided another set of fitting formulae for 2D 
X-ray and/or SZ surface brightness density profiles of galaxy clusters. 
Unlike the 3D formulae, the 2D formulae are expressed in terms of not
only $\mu$ but also the orientation angles of the line-of-sight in the
halo principal axes. In other words, we have found a link of the 2D
X-ray and/or SZ observables to the shapes and the orientations of the 
dark matter halos.

We have proposed a numerical algorithm based on our model to determine
the halo eccentricities and orientation angles from the observed X-ray
and SZ surface brightness density profiles of galaxy clusters.  Ideally
we have to apply our algorithm to real observation data, but
that is beyond the scope of the present paper since it should involve
careful treatment of the data image analysis together with various
observational systematic effects. Thus we have decided to test the
reconstruction algorithm against simple numerical profiles. We found
that the algorithm can reconstruct the halo eccentricities fairly
accurately if the halo eccentricities are greater than $0.3$.  On the
other hand, it turns out that the reconstruction of the halo orientation
angles are not always accurate, showing large scatters.

Finally, we conclude that our hydrostatic-equilibrium model for 
intra-cluster gas in triaxial dark halos makes it possible in principle 
to reconstruct the 3D structures of the dark halos from the X-ray and/or 
SZ cluster maps,  as long as the intra-cluster gas and the dark halos
are well approximated to be in hydrostatic equilibrium and the triaxial 
ellipsoids with a constant axis ratio, respectively.
We plan to test our model against real observational data, and hope to
report the results elsewhere in the future.

\acknowledgments 

We thank the anonymous referee for his/her careful reading 
of the manuscript and several constructive criticisms.  We are grateful 
to Masamune Oguri for useful discussions.  J. L. acknowledges gratefully 
the research grant of the JSPS (Japan Society of Promotion of Science) 
fellowship. This research was supported in part by the Grant-in-Aid 
for Scientific Research of JSPS (12640231).

\clearpage

\clearpage
\begin{table}
\caption{ Best-fit polynomial coefficients for the profiles of 
halo gravitational potentials}
\begin{center}
\begin{tabular}{lccccccccc}\hline \hline
   & \multicolumn{3}{c}{Isothermal}&&\multicolumn{3}{c}{Polytropic} \\
($\mu$)& $\beta$ & $q$   & $\eta$&& $\beta$ &  $q$  & $\eta$ \\ \hline
~0    & 1.329   & 0.426 & 0.146 && 0.780   & 0.617 & -0.002  \\ 
~1    & 0.118   & -0.246& 0.036 && 0.470   &-0.487 & 0.133   \\ 
~2    &-0.010   & 0.044 & 0.009 &&-0.087   & 0.136 &-0.013    \\ \hline
\end{tabular}
\end{center}
\label{table:3d_coef}
\end{table}

\clearpage
\begin{table}
\caption{ Best-fit polynomial coefficients for the profiles of the surface 
brightness in the isothermal case.}
\begin{center}
\begin{tabular}{lccccccccc}\hline \hline
   & \multicolumn{3}{c}{X-ray}&&\multicolumn{3}{c}{SZ} \\
($\mu$~ $\kappa$)& $\zeta$ &  $t$   & $\omega$ &&$\zeta$ &  $t$  & $\omega$ \\ \hline
~0~0   &  3.228  & -1.127 & -0.121  && -3.140 &  0.447 & 0.014   \\ 
~0~1   & -0.206  &  0.368 &  0.044  &&  1.133 & -0.087 & -0.005   \\ 
~0~2   &  0.008  &  0.026 & -0.002  && -0.058 &  0.018 & 0.001    \\ \hline
~1~0   & -0.850  &  3.749 & -0.010  &&  3.874 & -0.298 & 0.056    \\ 
~1~1   &  1.354  & -1.314 & -0.032  && -1.786 &  0.101 & -0.015   \\ 
~1~2   & -0.126  &  0.037 &  0.003  &&  0.166 & -0.027 & 0.000     \\ \hline
~2~0   & -12.23  & -2.184 &  0.158  &&  0.297 & -0.003 & -0.042    \\ 
~2~1   &  2.253  & 0.840  & -0.025  &&  0.342 & -0.015 & 0.014    \\ 
~2~2   & -0.082  & -0.044 &  0.000  && -0.064 &  0.011 & -0.001    \\ 
\hline 
\end{tabular}
\end{center}
\label{table:iso_coef}
\end{table}

\clearpage
\begin{table}
\caption{ Best-fit polynomial coefficients for the profiles of the surface 
brightness in the polytropic case with $\gamma = 1.15$.}
\begin{center}
\begin{tabular}{lccccccccc}\hline \hline
   & \multicolumn{3}{c}{X-ray}&&\multicolumn{3}{c}{SZ} \\
($\mu$~ $\kappa_{p}$)& $\zeta$ &  $t$   & $\omega$ &&$\zeta$ &  $t$  & 
$\omega$ 
\\ \hline
~0~0   &  2.261  & -0.943 & -0.257   && -3.667 & 2.542 &  0.101 \\ 
~0~1   & -2.689  &  0.753 &  1.027   &&  16.32 &-7.786 & -0.508  \\ 
~0~2   &  1.609  &  6.874 & -1.096   && -19.30 & 7.135 &  0.885  \\ 
~0~3   & -0.526  &  1.328 &  0.323   &&  7.156 & 2.345 & -0.483  \\ \hline
~1~0   & -0.049  & 30.48  &  0.371   && -3.390 &-19.12 &  0.424   \\ 
~1~1   &  13.86  &-112.8  & -2.301   && -0.685 & 77.61 & -1.296  \\ 
~1~2   & -27.91  & 123.3  &  3.722   &&  19.05 &-100.4 &  0.845 \\ 
~1~3   &  14.97  &-54.25  & -1.647   && -14.10 & 32.19 &  0.145  \\ \hline
~2~0   & -61.22  &-67.95  &  0.836   &&  36.73 & 34.75 & -1.279 \\ 
~2~1   &  197.2  & 268.9  & -1.902   && -125.8 &-147.3 &  4.857  \\ 
~2~2   & -199.1  &-331.6  &  0.485   &&  129.1 & 201.9 & -5.537  \\ 
~2~3   &  63.66  & 139.9  &  0.440   && -39.35 &-79.83 &  1.742  \\ \hline
~3~0   &  56.23  & 38.90  & -0.874   && -28.90 &-18.29 &  0.806  \\ 
~3~1   & -201.0  &-157.9  &  2.863   &&  107.2 & 78.73 & -3.186  \\ 
~3~2   &  227.3  & 202.3  & -2.703   && -123.9 &-110.4 &  3.918 \\ 
~3~3   & -82.35  &-86.20  &  0.743   &&  44.58 & 46.37 & -1.428  \\ 
\hline 
\end{tabular}
\end{center}
\label{table:poly_coef}
\end{table}


\clearpage

\clearpage

\clearpage

\clearpage

\clearpage

\clearpage

\clearpage

\end{document}